# Dynamic Model of the Price Dispersion of Homogeneous Goods

**Joachim Kaldasch[1*]**

*[1]EBC Hochschule Berlin, Alexanderplatz 1, 10178 Berlin, Germany.*

***Author's contribution***

*The sole author designed, analyzed and interpreted and prepared the manuscript.*

## ABSTRACT

Presented is an analytic microeconomic model of the temporal price dispersion of homogeneous goods in polypoly markets. This new approach is based on the idea that the price dispersion has its origin in the dynamics of the purchase process. The price dispersion is determined by the chance that demanded and supplied product units meet in a given price interval. It can be characterized by a fat-tailed Laplace distribution for short and by a lognormal distribution for long time horizons. Taking random temporal variations of demanded and supplied units into account both the mean price and also the standard deviation of the price dispersion are governed by a lognormal distribution. A comparison with empirical investigations confirms the model statements.

**Corresponding author: E-mail: joachim.kaldasch@international-business-school.de;*

## 1. INTRODUCTION

Based on the idea that both the demand and the supply side of a market can take advantage of arbitrage opportunities the classic theory suggests that a homogenous good must sell for the same price, known as "the law of one price" [1]. However, even for homogeneous goods, empirical investigations show the existence of price dispersion [2]. Economists give four popular explanations for the origin of the price dispersion in product markets of homogeneous goods: amenities, heterogeneous costs, intertemporal price discrimination and search frictions. The first explanation suggests that identical goods sell at different prices because they are bundled with different amenities in different transactions [3]. The second states that firms at different locations have different costs causing prices to vary for similar goods [4]. Time dependent fluctuations of the price in order to satisfy different consumer groups [5- 8] and the limited ability of buyers to search the entire market [9-11] are other economic explanations for the price dispersion. Previous models of spatial and temporal price dispersion separate between different consumer groups (e.g. informed and uninformed consumers) and establish a price distribution for profit maximizing competing sellers [10,12].

The presented dynamic model, however, suggests that the price dispersion has its origin in the dynamics of the purchase process. This dynamics is governed by the chance that supplied and demanded product units meet in a given price interval [13,14]. This novel approach concludes that independent of the economic explanation of price fluctuations, the price dispersion of homogeneous goods in polypoly markets must have the form of a fat-tailed distribution for short time horizons. This result is in agreement with empirical data and not predicted by previous theories of temporal price dispersion [15].

In order to derive the price dispersion the paper is organized as follows. Starting with fundamental relations governing the dynamics of the purchase process a quasi-static price dispersion is established. It takes advantage from the price dependent configuration of demanded and supplied units in the purchase process. By including time-dependent variations of supply and demand a Walrus equation describing the evolution of the mean price can be found. For the case of random demand and supply variations a lognormal mean price distribution is derived. Both distributions determine the unconditional price dispersion of a homogeneous good for short and long time horizons. The theory is compared with empirical investigations of an extensive study of homogeneous consumer goods over short time horizons performed by Kaplan and Menzio [15] followed by a conclusion.

## 2. THE MODEL

The presented model is established for a polypoly market of a homogeneous good with a large number of independent retailers (suppliers) *N(t)* and buyers *n(t)* at time step *t*. The dynamics of the polypoly market is formulated in terms of four presumptions.

### 2.1 Presumption 1

Purchase events in a market of homogeneous goods are the result of the conjuncture of demanded and supplied product units. Indicating the total number of demanded (desired) units by $\tilde{x}(t)$ and the total number of supplied (available) units by $\tilde{z}(t)$, purchase events must disappear if one of the variables vanishes. Hence, the total unit sales $\tilde{y}(t)$ counting the number of purchase events per unit time can be written up to the first order as a product of both variables [13]:

$$\tilde{y}(t) \cong \eta \tilde{z}(t)\tilde{x}(t) \qquad (1)$$

with the unknown rate $\eta$. This rate characterizes the mean frequency by which the meeting of demanded and supplied units generates successful purchase events (transactions). Since $\tilde{x}(t), \tilde{z}(t), \tilde{y}(t) \geq 0$, we demand that also $\eta \geq 0$.

### 2.2 Presumption 2

The demand and supply dynamics of the total physical flow of product units can be formulated in terms of conservation relations of the form[1]:

$$\frac{d\tilde{x}(t)}{dt} = \tilde{d}(t) - \tilde{y}(t); \qquad \frac{d\tilde{z}(t)}{dt} = \tilde{s}(t) - \tilde{y}(t) \qquad (2)$$

*[1] The total number of demanded and supplied units $\tilde{x}(t)$ and $\tilde{z}(t)$ are regarded to be scaled by a large number, such that these variables can be treated as real numbers. Therefore the time evolution is written in terms of differential equations.*

The first relation states that the total number of units demanded by buyers of the good increases with the total demand rate $\tilde{d}(t)$ and decreases by the purchase of product units with the total unit sales rate $\tilde{y}(t)$ . The second relation suggests that the total number of supplied units increases with the total supply flow $\tilde{s}(t)$ and decreases with the total unit sales rate $\tilde{y}(t)$ .

The presented theory of the price dispersion is based on a generalization of Eq.(1). For this purpose we introduce price dependent unit sales *y(t,p)*. For a homogenous good *y(t,p)* can be gained by accumulating the unit sales of individual sellers in a given price interval *p* and *p+dp* by:

$$y(t,p)=\sum_{k=1}^{N(t)} y_k(t,p) \tag{3}$$

The total unit sales can be obtained by integrating over the price dependent unit sales:

$$\tilde{y}(t)=\int_0^{\infty} y(t,p)dp \tag{4}$$

### 2.3 Presumption 3

The key idea establishing the price dispersion of a homogeneous good is that Eq.(1) is valid also for the meeting of demanded and supplied units in an arbitrary price interval *p* and *p+dp*. That means, *y(t,p)* must be zero if either the number of supplied units *z(t,p)* or the number of demanded units *x(t,p)* disappears in this price interval. Hence, the price dependent unit sales can be written up to the first order as proportional to the product:

$$y(t,p)\cong \eta z(t,p)x(t,p) \tag{5}$$

where the meeting rate *η* is considered to be price independent.

### 2.4 Presumption 4

Since all units of the good are equivalent buyers prefer to purchase units for the lowest available price. The willingness to purchase product units increases therefore with decreasing price. In order to make profit we further presume that the supply side will offer product units only for prices $p>p_m$, where the minimum price is strictly positive $p_m \geq 0$. The willingness to sell product units by sellers increases with increasing price.

The price dispersion can be given by the probability density *P(t,p)* of sold units:

$$P(t,p)=\frac{y(t,p)}{\tilde{y}(t)} \tag{6}$$

while the price dependent cumulative distribution has the form:

$$F(t,p)=\int_{p_m}^{p} P(t,p')dp' \tag{7}$$

The mean price *μ* of the good is determined by:

$$\mu(t)=\int_{p_m}^{\infty} P(t,p)pdp \tag{8}$$

The same procedure generating the price dependent unit sales *y(p,t)* can be also applied to establish the price dependent numbers of demanded units *x(p,t)* and supplied units *z(p,t)*. They are related to the total numbers by:

$$\tilde{z}(t)=\int_{p_m}^{\infty} z(t,p)dp; \qquad \tilde{x}(t)=\int_{p_m}^{\infty} x(t,p)dp \tag{9}$$

The corresponding probability density functions have the form:

$$P_z(t,p)=\frac{1}{\tilde{z}(t)}z(t,p); \qquad P_x(t,p)=\frac{1}{\tilde{x}(t)}x(t,p) \tag{10}$$

while the cumulative distributions are given by:

$$F_z(t,p)=\int_{p_m}^{p} P_z(t,p')dp'; \qquad F_x(t,p)=\int_{p_m}^{p} P_x(t,p')dp' \tag{11}$$

In order to characterize the temporal price dispersion we treat the time dependence of the total number of demanded and supplied as consisting of two parts. For a given time interval *Δt* they can be written as the sum of a constant mean part indicated by $\tilde{x}_0$ , $\tilde{z}_0$ and small time dependent variations denoted by *δx(t)*, *δz(t)*:

$$\tilde{x}(t)=\tilde{x}_0+\delta x(t); \qquad \tilde{z}(t)=\tilde{z}_0+\delta z(t) \tag{12}$$

The derivation of the price dispersion consists of three steps. First a quasi-static price dispersion $P_y(p)$ is derived for a constant number of demanded and supplied units $\tilde{x}_0$ and $\tilde{z}_0$. In the second step time dependent variations are taken into account as perturbations of $P_y(p)$. It turns out that they cause fluctuations of the mean price. A mean price distribution can be established for random variations of $\delta x(t)$ and $\delta z(t)$. In the third step the unconditional price dispersion $P(p)$ is determined by taking both effects into account.

### 2.5 The Quasi-static Price Dispersion

Under the condition that sufficient purchase events take place during $\Delta t$ the third presumption suggests that a quasi-static price distribution $P_y(p)$ can be established from the condition that the number of purchase events is determined by the chance that demanded and supplied units meet in a given price interval $p$ and $p+dp$. The fourth presumption further suggests the presence of a minimum price $p_m$. Buyers represent a number of demanded units $x(p)$ that must have its maximum $\tilde{x}_0$ at $p_m$ and disappears for $p\rightarrow\infty$. The number of demanded units $x(p)$ for buyers who wants to pay a price $p$ is determined by the accumulated number of units up to this price. It can be given in terms of the price dispersion of demanded units $P_x(p)$ by:

$$x(p) = \tilde{x}_0 - \tilde{x}_0 \int_{p_m}^{p} P_x(p')dp' = \tilde{x}_0\left(1 - F_x(p)\right) \quad (13)$$

Concerning the sellers, the fourth presumption suggests that the price dependent number of available units $z(p)$ increases with an increasing price and approaches its maximum $\tilde{z}_0$ for $p\rightarrow\infty$. The number of available units $z(p)$ is determined by the accumulation of the number of supplied units up to price $p$ and can be written as:

$$z(p) = \tilde{z}_0 \int_{p_m}^{p} P_z(p')dp' = \tilde{z}_0 F_z(p) \quad (14)$$

The price dependent unit sales Eq.(5) become with Eq.(13) and Eq.(14):

$$y(p) = \eta \tilde{z}_0 F_z(p) \tilde{x}_0 \left(1 - F_x(p)\right) \quad (15)$$

Applying Eq.(4) leads to the total unit sales:

$$\tilde{y} = \eta \tilde{x}_0 \tilde{z}_0 \sigma \quad (16)$$

with:

$$\sigma = \int_{p_m}^{\infty} F_z(p)\left(1 - F_x(p)\right)dp \quad (17)$$

The price distribution Eq.(6) turns with Eq.(15) and Eq.(16) into:

$$P_y(p) = \frac{1}{\sigma} F_z(p)\left(1 - F_x(p)\right) \quad (18)$$

Since $F_z(p)$ characterizes the probability of finding supplied units and 1-$F_x(p)$ the probability of finding demanded units, their product expresses the chance that demanded and supplied units meet at price $p$.

A monotone decreasing and a monotone increasing function show a single interception point $p^*$ at which:

$$x(p^*) = z(p^*) \quad (19)$$

For a price $p<p^*$, the number of desired units increases the number of available units $x(p)>z(p)$. Therefore the unit sales are limited by the number of available units $z(p)$ and the relative abundance of purchase events in the price interval $p$ and $p+dp$ is approximately equal to the relative abundance of available units. Hence $P_y(p)\approx P_z(p)$. For small prices the chance to find demanded units can be given by $1-F_x(p)\approx 1$. Therefore the price distribution Eq.(18) has for $p\le p^*$ the form:

$$P_y(p) \cong \frac{1}{\sigma} F_z(p) = \frac{1}{\sigma}\int_{p_m}^{p} P_z(p')dp' \cong \frac{1}{\sigma}\int_{p_m}^{p} P_y(p')dp' = \frac{1}{\sigma} F_y(p) \quad (20)$$

From this relation follows for $p\le p^*$:

$$P_y(p) = \frac{dF_y(p)}{dp} \cong \frac{1}{\sigma} F_y(p) \quad (21)$$

and thus:

$$F_y(p) = C_1 e^{\frac{p}{\sigma} - C_2} \tag{22}$$

while for convenience the integration constant is written as the product of two constants $C_1$ and $C_2$.

On the other hand for $p>p^*$ it can be argued that the number of available units increases the number of desired units $x(p)<z(p)$. Purchase events are therefore limited by the number of desired units and the relative abundance of purchase events is approximately equal to the relative abundance of desired units $P_y(p) \approx P_x(p)$. For large prices we can further approximate the chance to find supplied units by $F_z(p) \approx 1$ and Eq.(18) becomes for $p \geq p^*$:

$$P_y(p) \cong \frac{1}{\sigma}\left(1 - F_x(p)\right) = \frac{1}{\sigma}\left(1 - \int_{p_m}^{p} P_x(p')dp'\right) \cong \frac{1}{\sigma}\left(1 - \int_{p_m}^{p} P_y(p')dp'\right) = \frac{1}{\sigma}\left(1 - F_y(p)\right) \tag{23}$$

We get for $p \geq p^*$:

$$P_y(p) = \frac{dF_y(p)}{dp} \cong \frac{1}{\sigma}\left(1 - F_y(p)\right) \tag{24}$$

and hence:

$$F_y(p) = 1 - C_1 e^{\frac{-p}{\sigma} - C_3} \tag{25}$$

with the integration constant $C_3$. We demand for $p=p^*$ that:

$$C_1 e^{\frac{p^*}{\sigma} - C_2} = 1 - C_1 e^{\frac{-p^*}{\sigma} - C_3} \tag{26}$$

This relation can be satisfied by setting $C_1=1/2$, $C_2=p^*/\sigma$ and $C_3=-p^*/\sigma$. The cumulated distribution function has in the presented approximation the form:

$$F_y(p) \cong \begin{cases} \frac{1}{2} e^{\frac{p-\mu}{\sigma}} & p \leq \mu \\ 1 - \frac{1}{2} e^{-\frac{p-\mu}{\sigma}} & p \geq \mu \end{cases} \tag{27}$$

where we used that $p^*=\mu$ for a symmetric distribution. The quasi-static price distribution can be given by a fat-tailed Laplace distribution with the probability density function:

$$P_y(p) = \frac{1}{2\sigma} e^{-\frac{|p-\mu|}{\sigma}} \tag{28}$$

The parameter $\sigma$ is determined by the condition:

$$\sigma = \int_{\mu-\Delta}^{\mu} \frac{1}{2} e^{\frac{p-\mu}{\sigma}} dp + \int_{\mu}^{\infty} \frac{1}{2} e^{-\frac{p-\mu}{\sigma}} dp = \sigma\left(1 - \frac{1}{2} e^{-\frac{\Delta}{\sigma}}\right) \tag{29}$$

For a sufficiently small price difference $\Delta=\mu-p_m$ the exponential function can be approximated by the linear function:[2]

$$e^{-\frac{\Delta}{\sigma}} \sim 1 - \frac{\Delta}{\sigma} \tag{30}$$

and we obtain from Eq.(29):

$$\sigma \cong (\mu - \mu_m) \tag{31}$$

with $p_m=\mu_m$. The variance of the price distribution is given by the integral:

$$Var(P_y(p)) = \int_{\mu-\Delta}^{\mu} \frac{1}{2\sigma} e^{\frac{p-\mu}{\sigma}} (p-\mu)^2 dp + \int_{\mu}^{\infty} \frac{1}{2\sigma} e^{-\frac{p-\mu}{\sigma}} (p-\mu)^2 dp \tag{32}$$

It turns with Eq.(30) into:

$$Var(P_y(p)) \cong 2\sigma^2 \tag{33}$$

The standard deviation of the price dispersion is $\sqrt{2}\sigma$. Eq.(31) suggests that the standard deviation is a function of the mean price. Hence, the price dispersion becomes a very narrow peak for $\mu \approx \mu_m$. With Eq.(20) and Eq.(23) we can further approximate Eq.(13) and Eq.(14) by:

$$x(p) \cong \tilde{x}_0(1 - F_y(p)) \quad z(p) \cong \tilde{z}_0 F_y(p) \tag{34}$$

From the interception point Eq.(19) follows for the total number of supplied and demanded units:

[2] *While the empirical price dispersion $P_y(p)$ is bounded by $\mu_m$, the continuous description Eq.(28) is nonzero except at infinity. The approximation Eq.(30) is applied in order to take the limitation by the floor price into account. It is the origin of the mean price dependence of the standard deviation. The difference between the exponential and the linear function at $\mu_m$ expresses the error that is made by applying the continuous model.*

$$\tilde{x}_0 = \tilde{z}_0 \quad (35)$$

### 2.6 The Walrus Equation

The derivation of the quasi-static price dispersion $P_y(p)$ neglects the impact of time dependent variations $\delta x(t)$ and $\delta z(t)$. Eq.(31) suggests that the only free variable of the quasi-static price dispersion Eq.(28) is $\mu$. Therefore variations in $\delta x(t)$ and $\delta z(t)$ can be taken into account by time dependent variations of the mean price. In order to establish a relation for $\mu(t)$, the interception point Eq.(19) at time step $t$ is written as:

$$x_{old}(\mu_{old}) = z_{old}(\mu_{old}) \quad (36)$$

and the new interception point at $t+\Delta t$ as:

$$x_{new}(\mu_{new}) = z_{new}(\mu_{new}) \quad (37)$$

The new mean price is shifted in relation to the old mean price by:

$$\mu_{new} = \mu_{old} + \delta\mu \quad (38)$$

The new interception point can be written as a perturbation of the old interception point by:

$$\begin{aligned} x_{new}(\mu_{new}) &= x_{old}(\mu_{new}) + \delta x; \\ z_{new}(\mu_{new}) &= z_{old}(\mu_{new}) + \delta z \end{aligned} \quad (39)$$

while:

$$\begin{aligned} x_{old}(\mu_{new}) &= x_{old}(\mu_{old}) + \left.\frac{dx_{old}(p)}{dp}\right|_{\mu_{old}} \delta\mu; \\ z_{old}(\mu_{new}) &= z_{old}(\mu_{old}) + \left.\frac{dz_{old}(p)}{dp}\right|_{\mu_{old}} \delta\mu \end{aligned} \quad (40)$$

Applying Eq.(34) the price derivatives become:

$$\left.\frac{dx_{old}(p)}{dp}\right|_{\mu} = -\tilde{x}_0 P_y(\mu) = \frac{-\tilde{x}_0}{2\sigma} \quad (41)$$

and:

$$\left.\frac{dz_{old}(p)}{dp}\right|_{\mu} = \tilde{x}_0 P_y(\mu) = \frac{\tilde{x}_0}{2\sigma} \quad (42)$$

where we used Eq.(35). Inserting these relations in Eq.(37) the mean price change during a short time interval $\Delta t \rightarrow dt$ turns with Eq.(31) into:

$$\frac{1}{\mu - \mu_m}\frac{d\mu}{dt} = H\left(\frac{d\tilde{x}}{dt} - \frac{d\tilde{z}}{dt}\right) \quad (43)$$

with

$$H = \frac{1}{\tilde{x}_0} \quad (44)$$

Eq.(43) can be rewritten with Eq.(2) as:

$$\frac{1}{\mu(t) - \mu_m}\frac{d\mu(t)}{dt} = H(\tilde{d}(t) - \tilde{s}(t)) \quad (45)$$

This relation is the well-known Walrus equation [16]. It states that an excess total demand rate increases and an excess total supply rate decreases the mean price in time.

### 2.7 The Mean Price Distribution

In order to establish a dynamic model of the price dispersion we assume that both sides of the market generate random variations of the demand and supply rates such that:

$$\chi(t) = H(d(t) - s(t)) \quad (46)$$

can be treated as a fluctuating function. As a first approximation $\chi(t)$ is described as white noise with time average and time correlation function:

$$\begin{aligned} \langle \chi(t) \rangle &= 0 \\ \langle \chi(t), \chi(t') \rangle &= 2D\,\delta(t - t') \end{aligned} \quad (47)$$

where $D$ is a noise amplitude. The Walrus relation Eq.(45) turns in this case into a Langevin equation of the form:

$$\frac{d\omega(t)}{dt} = \chi(t)\omega(t) \quad (48)$$

with a shifted mean price:

$$\omega(t) = \mu(t) - \mu_m \quad (49)$$

The mean price $\omega(t)$ is governed in this approximation by a multiplicative stochastic process. The central limit theorem suggests that the mean price distribution can be given after sufficient time by a lognormal distribution of the form:

$$P_\omega(\omega) = \frac{1}{\sqrt{2\pi}\Omega\omega}\exp\left(-\frac{(\ln(\omega/\Gamma))^2}{2\Omega^2}\right) \quad (50)$$

where $\Gamma$ and $\Omega$ are free parameters.

Eq.(31) and Eq.(33) suggest that the standard deviation of the price dispersion $P_y(p)$ is proportional to $\omega$. Therefore the standard deviation must be also considered as a fluctuating variable. The parameter $\sigma$ is in this approximation governed by a lognormal distribution that can be obtained from Eq.(50) by changing variables:

$$P(\sigma) = \frac{1}{\sqrt{2\pi}\Omega\sigma}\exp\left(-\frac{(\ln(\sigma/\Gamma))^2}{2\Omega^2}\right) \quad (51)$$

### 2.8 The Price Dispersion

The model suggests that the price of a homogeneous good is determined by two random processes. On the one hand it is related to the chance that demanded and supplied units meet in a given price interval leading to the quasi-static price distribution $P_y(p)$. On the other hand demand and supply variations generate fluctuations of the mean price described by $P_\omega(\omega)$. In order to take both effects into account Eq.(6) can be interpreted as an unconditional price distribution of the form:

$$P(p) = \int_0^\infty P_y(p\,|\,\omega)P_\omega(\omega)d\omega \quad (52)$$

where the conditional price distribution $P_y(p|\omega)$ characterizes the probability density under the condition that the mean price is given by $\omega$.

We want to confine the discussion here to two extreme cases:

*i)*. In the first case the price distribution $P_y(p)$ is regarded to be located around the mean price. The conditional price distribution $P_y(p|\omega)$ can then be approximated by a Dirac-delta function of the form:

$$P_y(p\,|\,\omega) \approx \delta(p-\omega) \quad (53)$$

and the unconditional price distribution becomes:

$$P(p) = \int_0^\infty \delta(p-\omega)\frac{1}{\sqrt{2\pi}\Omega\omega}\exp\left(-\frac{(\ln(\omega)-\Gamma)^2}{2\Omega^2}\right)d\omega = \frac{1}{\sqrt{2\pi}\Omega p}\exp\left(-\frac{(\ln(p)-\Gamma)^2}{2\Omega^2}\right) \quad (54)$$

That means demand and supply variations govern the price distribution. After sufficient time the price dispersion has the form of a lognormal distribution.

*ii)*. In the second case we consider demand and supply variations as small. The mean price is in this case located around a constant $\mu_0$ and the price distribution Eq.(50) can be approximated by:

$$P_\omega(\mu) \approx \delta(\mu-\mu_0) \quad (55)$$

The price distribution becomes:

$$P(p) = \int_0^\infty \frac{1}{2\sigma}e^{-\frac{|p-\mu|}{\sigma}}\delta(\mu-\mu_0)d\mu = \frac{1}{2\sigma}e^{-\frac{|p-\mu_0|}{\sigma}} \quad (56)$$

In this case the price dispersion reduces to a Laplace distribution.

For short time periods the price dispersion of a homogenous good can be expected to be close to approximation *ii)*, because large variations of demand and supply rates are rather unlikely. Over long time periods, however, the price dispersion will be governed by approximation *i)*.

## 3. COMPARISION WITH EMPIRICAL DATA

For a comparison of the model with empirical data we want to confine here to homogeneous consumer goods studied over short time horizons. In this case the model suggests that the price dispersion can be approximated by the Laplace distribution Eq.(56). Normalizing this price dispersion with respect to the mean price $\mu_0$, the only free parameter is the parameter $\sigma$ which is up to a constant factor equivalent to the standard deviation.

An extensive investigation of price dispersions of homogeneous consumer goods was performed by Kaplan and Menzio [15]. The price dispersions were obtained by analysing data from the Kilts-Nielsen Consumer Panel Dataset (KNCP) [3]. Households in this panel provide information about each of their shopping trips using a Universal Product Code (UPC) scanning device provided by Nielsen. The panel tracks the shopping behaviour of approximately 50,000 households over the period 2004 to 2009 and contains price and quantity information for over 300 million transactions. The dataset covers over 1.4 million goods in 54 geographical markets, each of which roughly corresponds to a Metropolitan Statistical Area in the USA.

[3] *The Kilts-Nielsen Consumer Panel Data are supplied by the Kilts-Nielsen Data Center at the University of Chicago Booth School of Business. Information on data access and availability is available at http://research.chicagobooth.edu/nielsen*

The investigators aggregated the data into four different definitions of a good:

1. UPC: A good is the set of products with the same Universal Product Code (barcode).
2. Generic Brand Aggregation: According to this definition, a good is the set of products that share the same features, the same size and the same brand, but may have different UPC's. Since the KNCP assigns the same brand code to all generic brands (regardless of the retailer), this definition collects all generic brand products that are otherwise identical.
3. Brand Aggregation: According to this definition, a good is the set of products that share the same features and the same size, but may have different brands and different UPCs.
4. Brand and Size Aggregation: In this case a good is considered to be the set of products that share the same features but may have different sizes, different brands and different UPCs.

Displayed in the Figs. 1-4 are the empirical price dispersions adopted for each definition of a good by plotting the distribution across all investigated markets, goods and quarters. The investigators found (nearly) symmetric, leptokurtic price distributions of normalized prices *($\mu_0$=1).* Since the empirical distributions are fat-tailed the Laplace distribution can be used to fit the empirical price dispersions as shown by the solid lines in the figures.

In the first two definitions of a good the supplier dominates the market. We therefore conclude that deviations of the empirical data from the expected Laplace distribution around the centre in the Figs. 1-2 stem from the dominance of the main supplier in the homogeneous market.[4] The third and fourth definitions of a good come closer to the model assumptions. As can be seen in Figs. 3-4 the deviation from the Laplace distribution around the centre disappears. Note that in the Figs. 1-3 the price dispersion is symmetric around the mean price. The distribution is slightly shifted in Fig. 4 to lower values and deviates from a symmetric Laplacian for increasing prices. It indicates that approximation *ii)* is not completely valid and the mean price fluctuates considerably in this definition of a good.

The standard deviation of a homogenous good is related in the presented model by the difference between the actual mean price and the floor price $\mu_m$ (Eq.(31)). The empirically found increase of the standard deviation with increasing aggregation in the Figs. 1-4 is probably due to the increased variety of product versions. [5]

The presented theory predicts, though, that the mean price and hence the price standard deviation fluctuates in time governed by a lognormal distribution. Instead studying the time evolution of the standard deviation of a single good, we can alternatively think of an ensemble of homogeneous goods and investigate the distribution of their standard deviations. In other words, if sufficient homogeneous goods are available, the time averaged distribution of the standard deviation can be approximated by an ensemble average.

The empirical specification of the relative abundance of the standard deviation of all investigated markets, goods and quarters are displayed in Fig. 5. Under the condition that the time and ensemble average generates the same distribution, Eq.(51) suggests that the distribution of the standard deviation is lognormal starting at *$\omega$=0*. Setting out normalized prices, the lognormal distribution must therefore be shifted to negative values by the mean magnitude $\mu_m/\mu_0$. Applying a shifted lognormal distribution a good coincidence with the empirical data can be achieved as shown by the solid line in Fig. 5.[6]

*[4] The price distribution can be considered in this case as consisting of two contributions, one from the dynamics of dominant supplier and the other from independent retailers. The model is based on the purchase process and describes therefore merely the impact of the retailers on the price dispersion.*

*[5] For a single product version is $\mu_0$=$\mu_m$ and the standard deviation of the short term price dispersion disappears. With an increased product variety (number of sellers) the standard deviation increases, because the ensemble contains additional (e.g. local) fluctuations of demand and supply. This dependence is not further discussed here.*

*[6] Note that the same argumentation applies to the mean price. The short term distribution of the relative abundance of the absolute mean price of a sufficient number of homogeneous consumer goods must be lognormal.*

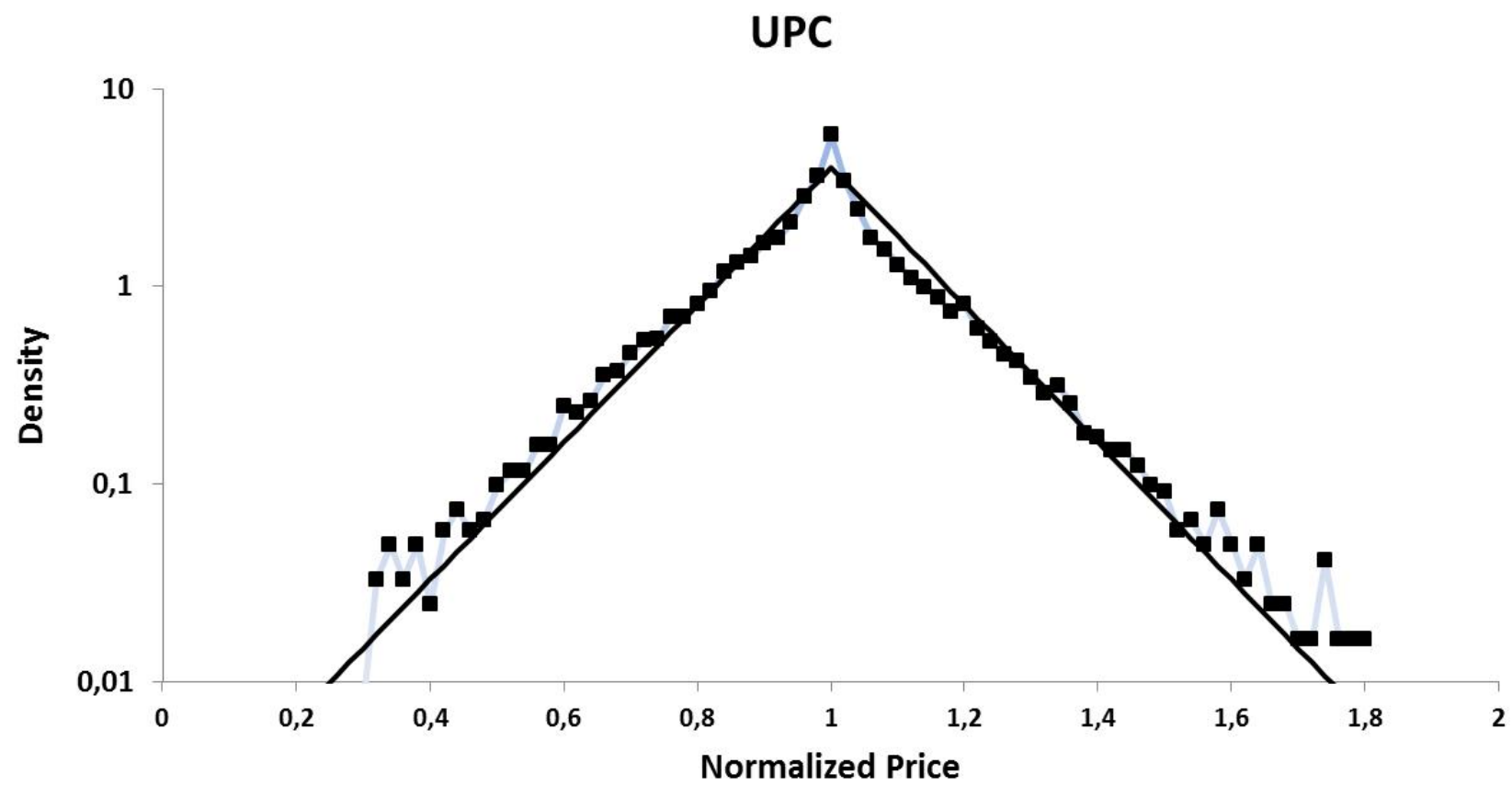

**Fig. 1. Displayed is the normalized price distribution (squares) for the first definition of a good [15]. The solid line indicates a Laplace distribution fit with *σ=0.125***

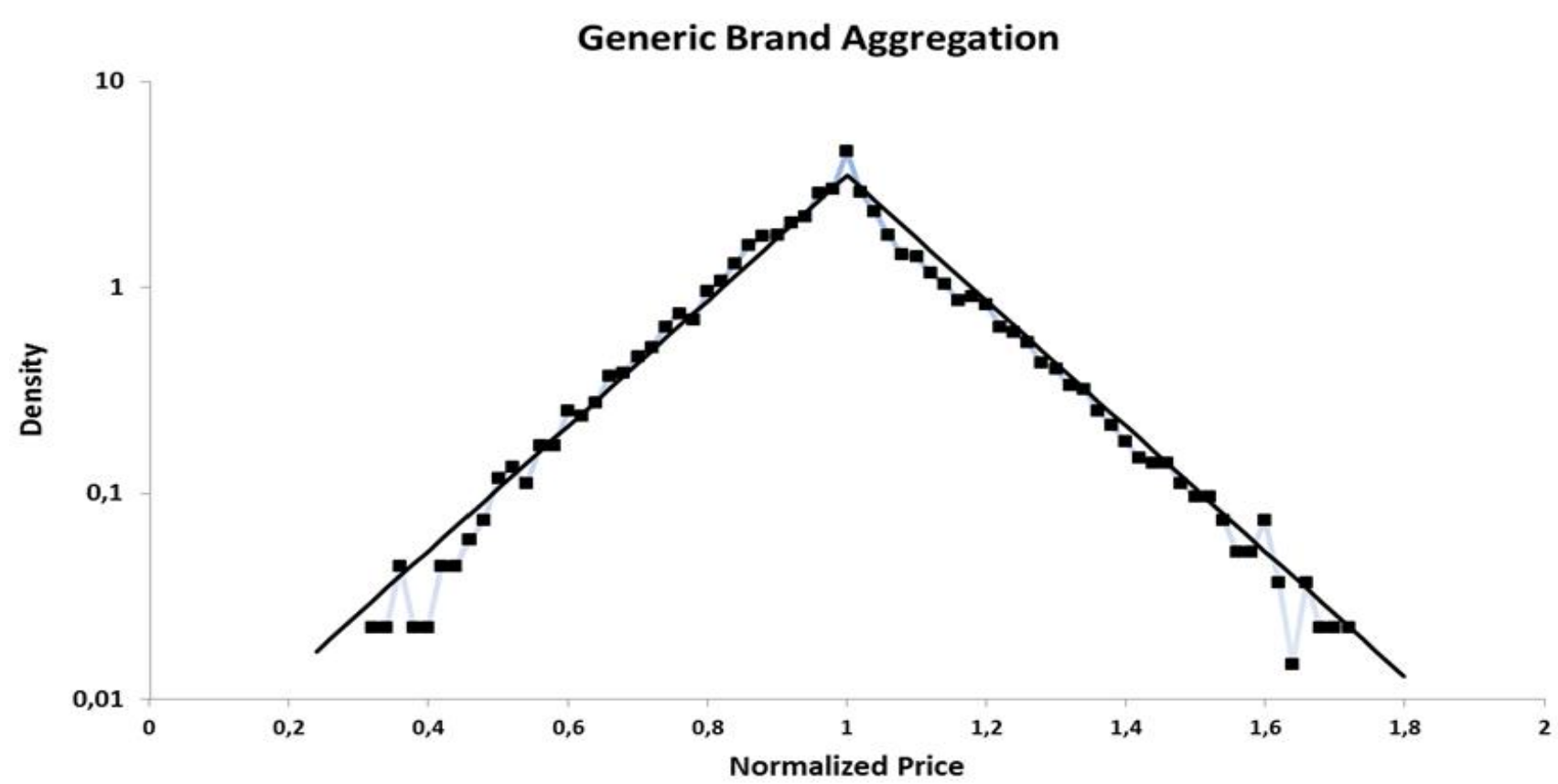

**Fig. 2. Displayed is the average price distribution (squares) for the second definition of a good [15]. The solid line is fit of a Laplace distribution with *σ=0.143***

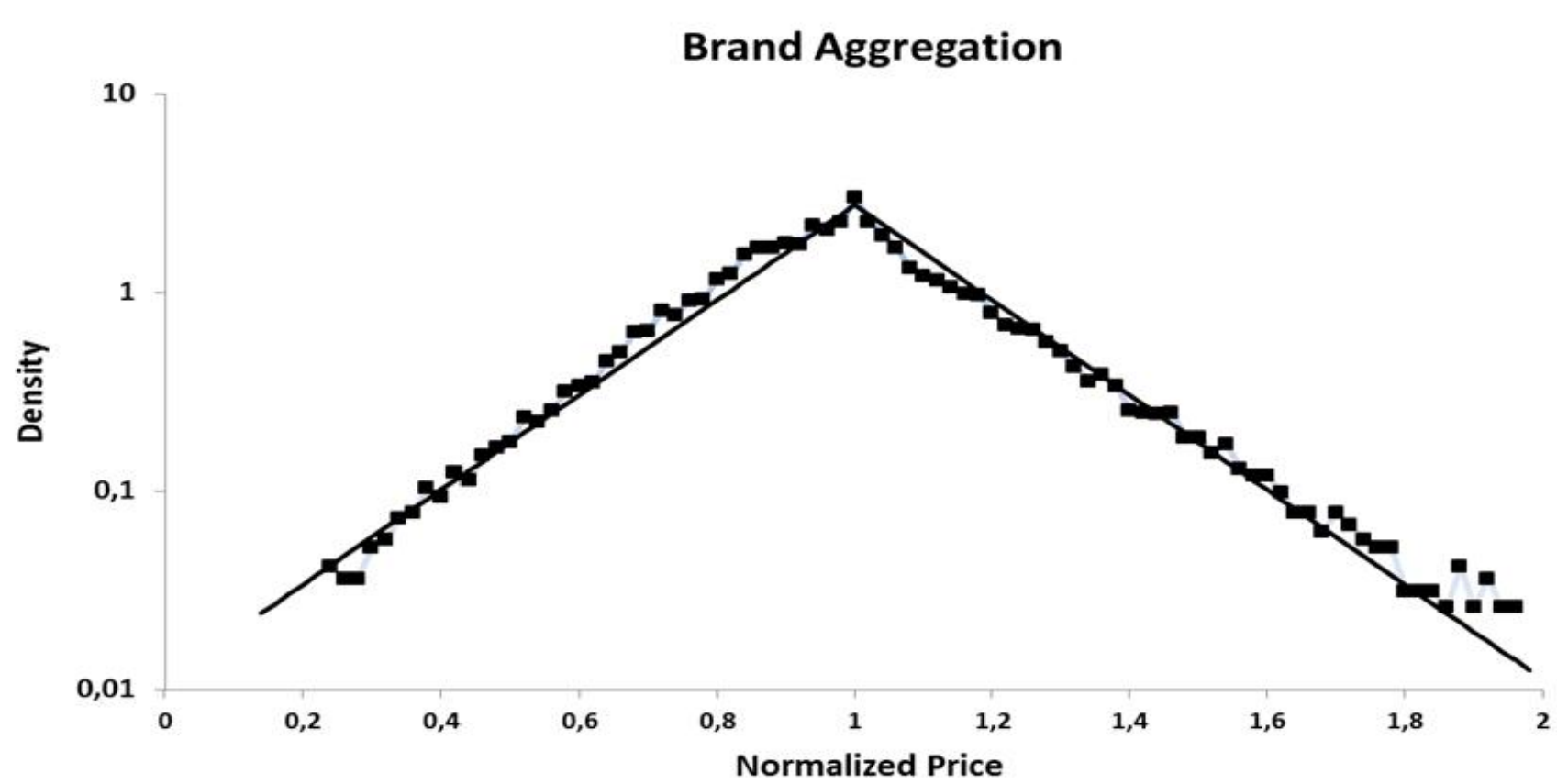

**Fig. 3. Displayed is the average price distribution (squares) for the third definition of a good [15]. The solid line is a fit of a Laplace distribution with *σ=0.145***

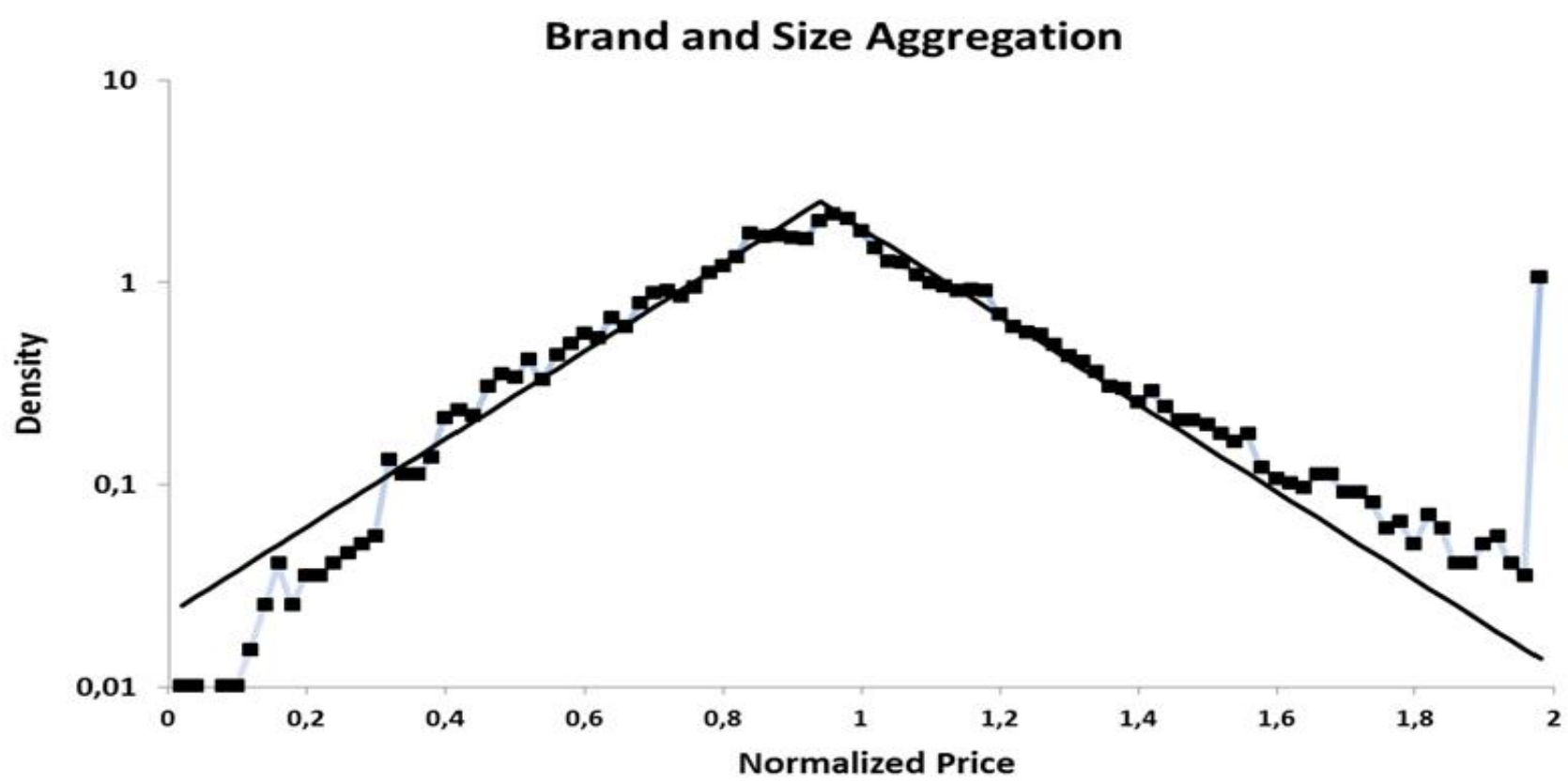

**Fig. 4. Displayed is the average price distribution (squares) for the fourth definition of a good [15]. The solid line indicates a Laplace distribution with $\sigma=0.2$ that is shifted by *0.06***

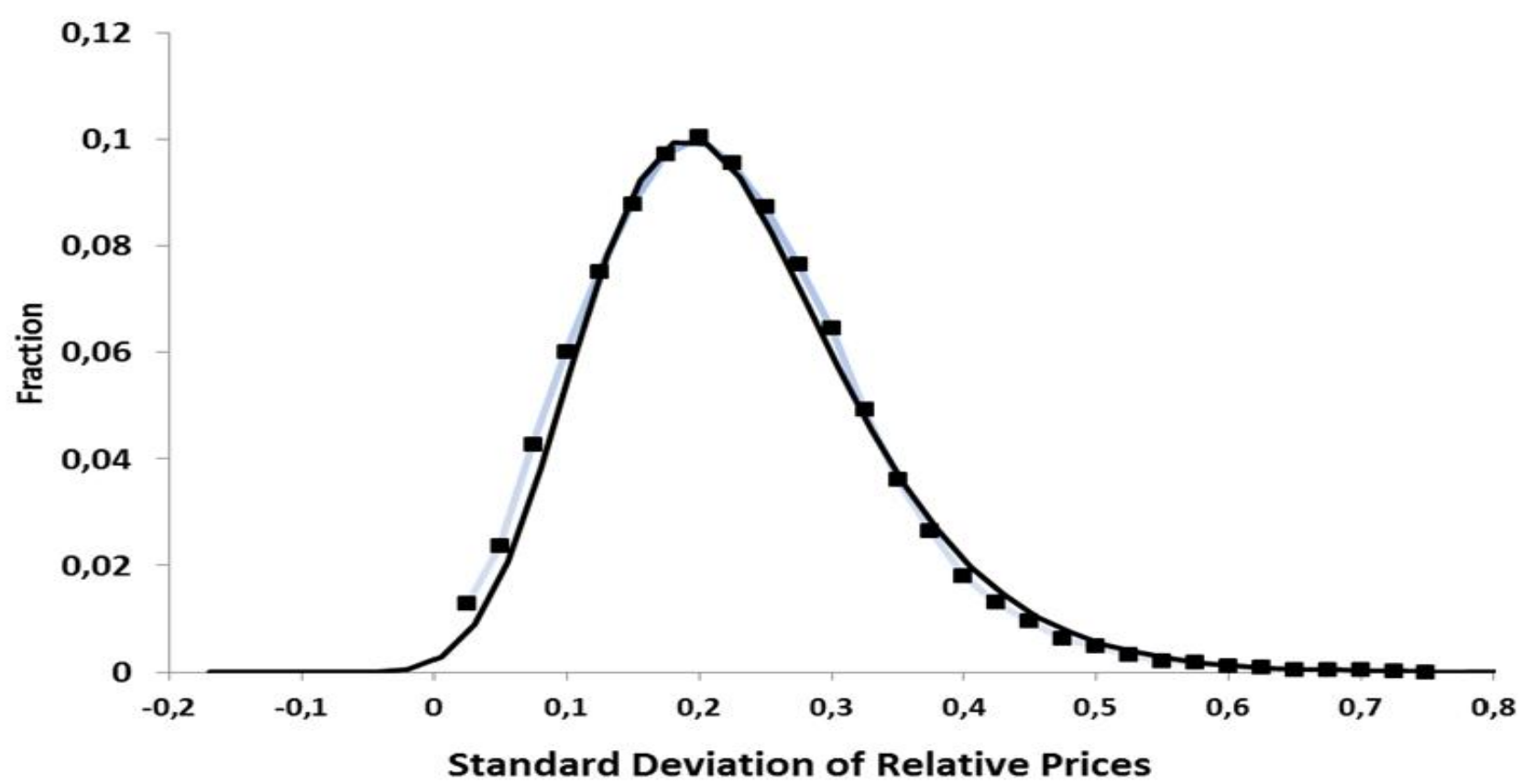

**Fig. 5. Displayed is the standard deviation of the relative price of the empirical data (squares) [15]. The solid line is a fit of a lognormal distribution with $\Gamma=0.41$, $\Omega=0.245$ shifted by $\mu_m/\mu_0=0.0245$**

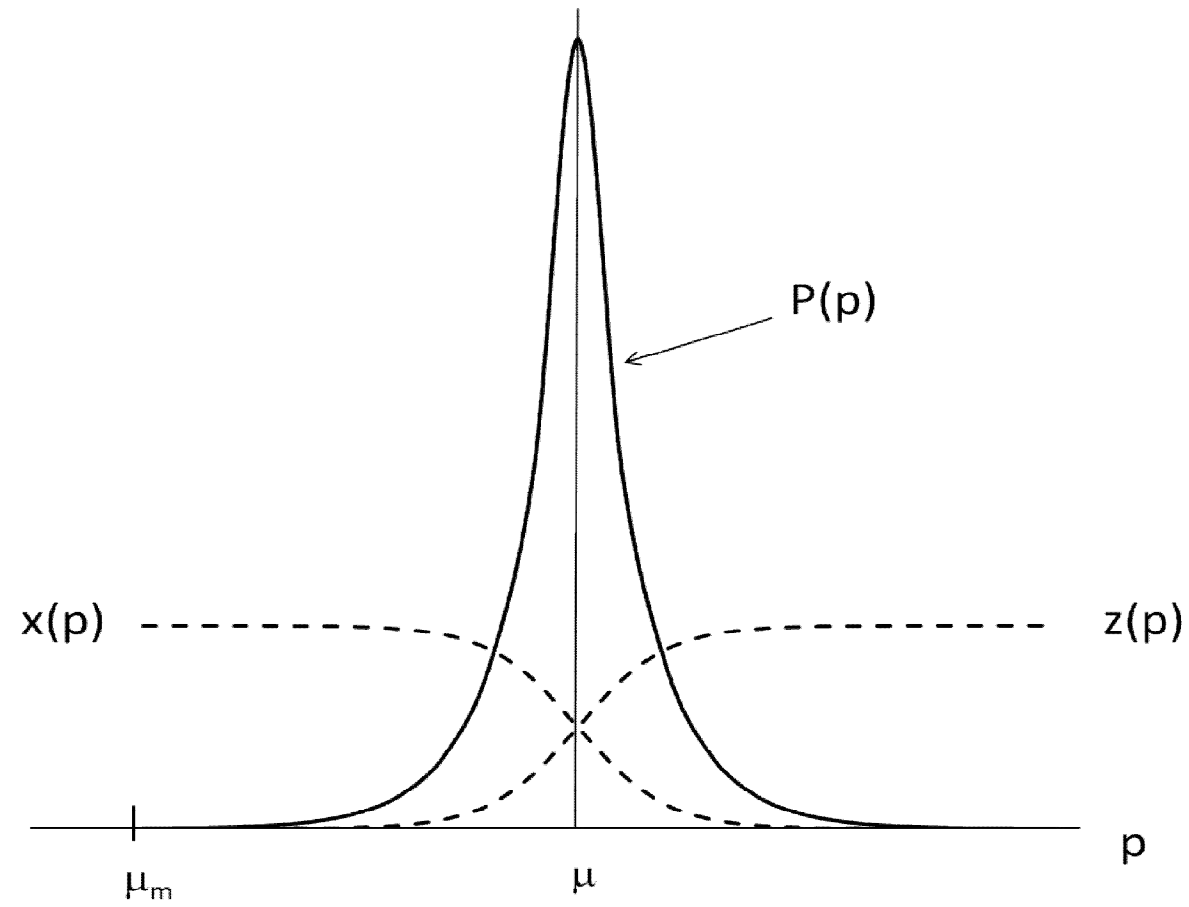

**Fig. 6. Schematically displayed are the price dependent distributions of demand units *x(p)* and supply units *z(p)* of a homogeneous good in a polypoly market (dashed lines). The corresponding short-term price dispersion *P(p)* is a Laplacian located around the mean price $\mu$ (solid line) bounded by the floor price $\mu_m$**

## 4. CONCLUSION

The paper establishes an analytic model of the price dispersion. Based on the assumption that the price dispersion is governed by the dynamics of the purchase process the presented approach predicts price dispersion of homogeneous goods. The theory suggests that the price dispersion can be approximated by a Laplace distribution for short and by a lognormal price distribution for long time horizons. A short term price dispersion *P(p)* is schematically displayed in Fig. 6 above together with the price dependent distributions of demanded and supplied units *x(p)* and *z(p)*. For $p<\mu$ the number of demanded units increases the number of supplied units and vice versa for $p>\mu$. The functions *x(p)* and *z(p)* intercept at mean price $\mu$. Following the classic approach the market is cleared at this price. However, the presented model takes the in- and outflows of product units into account. Since the chance that demanded and supplied units meet have its maximum at mean price, the price dispersion *P(p)* is located around $\mu$. It has a lower bond at the floor price $\mu_m \geq 0$. The standard deviation of the price dispersion is a function of the mean price. As a consequence the price distribution becomes a very narrow peak for $\mu \rightarrow \mu_m$.

The theory further suggests that the mean price is not constant but fluctuates in time due to demand and supply variations. For a homogeneous good the mean price is governed by a Walrus equation, suggesting that an excess demand increases and an excess supply decreases the mean price. For the case that demand and supply variations are subject to random fluctuations the distributions of the mean price and also of the standard deviation are governed by a lognormal distributions.

The theory is compared with empirical investigations of the short-term price dispersion of homogeneous consumer goods. As expected by the model the empirical data can be fitted with the fat-tailed Laplace distribution while the standard deviation is the only free parameter. The presented approach is, however, limited to a polypoly market. If there are large players in the market disturbing this condition, they may lead to additional contributions to the price dispersion. The comparison with empirical data suggests that deviations from the Laplace distribution around the centre may be caused by the dominance of suppliers. In agreement with the model the empirical data of the standard deviation of the price dispersion can be fitted with a shifted lognormal distribution.

The main result of the presented model is that the price dispersion of homogeneous goods in polypoly markets must exhibit the same stylized facts independent of the economic explanation of price variations.

## COMPETING INTERESTS

The author has declared that no competing interests exist.